# Observation of spontaneous spin-splitting in the band structure of an n-type zinc-blende ferromagnetic semiconductor

Le Duc Anh[1,2], Pham Nam Hai[3,4], and Masaaki Tanaka[1,4*]

[1]*Department of Electrical Engineering and Information Systems, The University of Tokyo, 7-3-1 Hongo, Bunkyo-ku, Tokyo 113-8656, Japan*
[2]*Institute of Engineering Innovation, Graduate School of Engineering, The University of Tokyo, 7-3-1 Hongo, Bunkyo-ku, Tokyo 113-8656, Japan*
[3]*Department of Physical Electronics, Tokyo Institute of Technology, 2-12-1 Ookayama, Meguro, Tokyo 152-0033, Japan*
[4]*Center for Spintronics Research Network (CSRN), The University of Tokyo, 7-3-1 Hongo, Bunkyo-ku, Tokyo 113-8656, Japan*

Large spin splitting in the conduction band (CB) and valence band (VB) of ferromagnetic semiconductors (FMSs), predicted by the influential mean-field Zener model[1,2] and assumed in many spintronic device proposals[3-8], has never been observed in the mainstream p-type Mn-doped FMSs[9-15]. Here using tunnelling spectroscopy in Esaki-diode structures, we report the observation of such a large spontaneous spin-splitting energy ($\Delta E$ = 31.7 - 50 meV) in the CB bottom of n-type FMS (In,Fe)As, which is surprising considering the very weak *s-d* exchange interaction reported in several zinc-blende (ZB) type semiconductors[16,17]. The mean-field Zener model also fails to explain consistently the ferromagnetism and the spin splitting energy $\Delta E$ of (In,Fe)As, because we found that the Curie temperature ($T_C$) values calculated using the observed $\Delta E$ are much lower than the experimental $T_C$ by a factor of 400. These results urge the need for a more sophisticated theory of FMSs. Furthermore, bias-dependent tunnelling anisotropic magnetoresistance (TAMR) reveals the magnetic anisotropy of each component of the (In,Fe)As band structure [CB, VB, and impurity band (IB)]. The results suggest that the energy **range of IB overlaps with the CB bottom or VB top, which may be important to understand the strong *s-d* exchange interaction in (In,Fe)As[18,19].**

Since the design of spintronic devices depends heavily on the knowledge of the band structure of the magnetic materials, the band structure of FMSs has been the central topic of active debates in the last decade. The most widely used theory of the ferromagnetism in FMSs, "mean-field Zener model"[1,2], which assumes that the Fermi level lies in the CB or VB of the host semiconductors, predicted that the *s,p-d* exchange interactions would induce spin-splitting in these bands, given by $\Delta E = (\alpha \text{ or } \beta)N_0 xS$, where $\alpha$ and $\beta$ are the *s-d* and *p-d* exchange integrals, respectively, $N_0$ is the cation site concentration, *x* is the magnetic atom concentration, and *S* is the spin angular momentum of each magnetic atom. This is the most desirable picture, because we can easily design the material properties and functions by using well-established band engineering of semiconductors[20]. However, recent experimental results in (Ga,Mn)As, which is widely recognized as a "canonical" FMS, have revealed that the Fermi level lies in the Mn-related impurity band (IB)[9-15], while the CB and VB of the host material remain nearly nonmagnetic (the "IB conduction picture")[12]. Small spontaneous spin-splitting *ΔE* (about 10 meV) in the VB has been observed only in p-type Mn-doped II-VI based FMS (Cd,Mn)Te quantum wells (QWs), which is much less practical due to the very low Curie temperature ($T_C < 4$ K)[21]. These results casted doubts on the validity of the mean-field Zener model as a universal model to describe and predict the magnetic properties of FMSs. The realization of FMSs with spin-split host semiconductor's band structure at higher temperature range, which will give a decisive impact on the development of semiconductor-based spintronic devices, also remains elusive.

In this article, we report on the observation of large *spontaneous* spin-splitting in the CB of an n-type ZB-type FMS, (In,Fe)As[22], studied by tunnelling spectroscopy using spin-Esaki-diode devices. The mean-field Zener model predicted no n-type ZB-type FMS because the *s-d* exchange interaction is generally very weak in these materials[16,17]; however, the electron-induced ferromagnetism has been found and unambiguously confirmed in (In,Fe)As and its $T_C$ can be controlled by changing the electron concentration or manipulating the electron wavefunction in (In,Fe)As QWs[18,22,23]. The Fermi level of n-type (In,Fe)As lies in the CB[18,19], which makes it free from the exotic IB conduction picture observed in (Ga,Mn)As. Furthermore, based on the mean-field Zener model, $N_0\alpha$ of (In,Fe)As was estimated to be about 2.8 eV ~ 4.5 eV[18,19,23], which implies that $T_C$ of the material would reach room temperature if a high enough electron density (~$10^{20}$ cm$^{-3}$) can be achieved. These $N_0\alpha$ values are unexpectedly large, thus raising questions on the conventional understanding of the *s-d* exchange interaction in FMSs and validity of the mean-field Zener model. Direct observation of the spin-dependent band structure of (In,Fe)As thus would give valuable insights into the ferromagnetism in this material and the physics of FMSs in general, as well as give an important basis for device applications using spin-dependent band engineering.

## RESULTS

**Preparation and *I-V* characteristics of (In,Fe)As-based Esaki diode devices**

The band structure of (In,Fe)As is studied by tunnelling spectroscopy using spin-Esaki-diode devices of n$^+$-(In,Fe)As / p$^+$-InAs. Figure 1a illustrates the diode devices used in this work, whose layered structure from the surface is 50 nm-thick n$^+$-type (In,Fe)As (with or without Be doping) / 5 nm-thick InAs / 250 nm-thick p$^+$-type InAs:Be (Be acceptor concentration $N_{Be} = 1 \times 10^{18}$ cm$^{-3}$) / p$^+$ type InAs (001) substrate. Two different samples (A and B) were prepared (see Methods) with different Fe concentrations (6% and 8%, respectively) and electron densities (by co-doping Be donors of $5 \times 10^{19}$ cm$^{-3}$ in the (In,Fe)As layer of device B), and consequently different $T_C$ (45 K and 65 K, respectively). Although the electron density $n$ of the (In,Fe)As layers in these two samples was not able to be measured by Hall effect measurements due to the parallel conduction in the p$^+$ type InAs substrates, the typical $n$ values for (In,Fe)As samples without and with Be co-doping are in the order of $1 \times 10^{18}$ and $1 \times 10^{19}$ cm$^{-3}$, respectively[19,22,23]. The two samples were then fabricated into mesa diodes with 200 μm in diameter (devices A and B) for *I-V* measurements. The bias polarity was defined so that in positive bias current flows from the p$^+$-type InAs substrate to the n$^+$-type (In,Fe)As layer.

Figure 1b illustrates the band profiles of these spin-Esaki diodes in different ranges of bias voltage *V*. We assume that (In,Fe)As has a bandgap energy $E_g$ and a spin-split CB (we do not take into account spin splitting of the VB of (In,Fe)As, because it is away from the Fermi level thus irrelevant to the present study). At *V* = 0, due to the heavy doping, the Fermi level (denoted by $E_F$) lies at $E_n$ above the CB bottom of n$^+$-(In,Fe)As and at $E_p$ below the VB top of p$^+$-InAs. The d*I*/d*V*-*V* curves of spin-Esaki diodes can be divided into three regions, as illustrated in Fig. 1b. At $V < e^{-1}(E_n + E_p)$ (*e* is the elementary charge), corresponding to region ① (tunnelling region), electrons tunnel from the CB of the n$^+$-(In,Fe)As to the VB of the p$^+$-InAs. Because the tunnelling conductance d*I*/d*V* is proportional to the product of the density of states (DOS) of the electrodes, we can probe the CB structure of the n$^+$-type (In,Fe)As from the d*I*/d*V*-*V* curves in this region. As illustrated in Fig. 1c, at low temperatures ($T < T_C$) or under external magnetic fields, the minority spin and majority spin CBs of (In,Fe)As split, which leads to "kink" structure in the d$I$/d$V$-$V$ curve in this region (pointed by two black arrows in Fig. 1c). Although the d$I$/d$V$ value also depends on the VB structure of the non-magnetic p$^+$-InAs, the VB of p$^+$-InAs (heavy hole and light hole bands) should show negligible change when applying a magnetic field up to 1 T, or when varying the temperature in the range around $T_C$ of the (In,Fe)As layers of the two devices (0 ~ 65 K). Thus we can elucidate the spin-dependent CB structure of (In,Fe)As by investigating the magnetic field and temperature dependence of the "kink" structure. When the bias voltage $V$ is equal to $e^{-1}(E_p + E_n)$, the (In,Fe)As CB bottom is lifted to the same energy as the p$^+$-InAs VB top and the current due to the CB-to-VB tunnelling is suppressed, leading to a negative differential resistance (NDR). At $e^{-1}(E_n + E_p) < V < e^{-1}E_g$, corresponding to region ② (bandgap region), the tunnelling of electrons into the bandgap is forbidden. However, there is possibly a small current caused by two factors; the thermionic current from electrons and holes thermally hopping and diffusing over the barrier at finite temperatures, and the tunnelling of electrons through the gap states which usually exist in heavy-doped semiconductors. This additional current may weaken or conceal the NDR characteristics. Finally at larger bias voltages $e^{-1}E_g < V$, corresponding to region ③, the occupied states in the CB (or VB) of n$^+$-(In,Fe)As reach the same energies as the unoccupied states in the CB (or VB) of p$^+$-InAs, then diffusive and drift currents start to flow as in normal pn-junction diodes. Thus we call this region ③ diffusion region.

Figures 1d and e show the d$I$/d$V$-$V$ curves measured in devices A and B, respectively, at 3.5 K. In both diode devices, we clearly see all the three regions, as indicated by colored areas in Figs. 1d,e: The border between the tunnelling region ① and the band-gap region ② is marked by a "bend" (indicated by blue triangles), observed around 60 mV in device

A and 180 mV in device B. Meanwhile, the starting of the diffusion region ③ is indicated by a sharp increase at around 0.42 V and 0.48 V in device A and B, respectively.

**Spin-dependent conduction band structure of (In,Fe)As**

We first focus on the tunnelling regions (region ①) in both devices, where electrons tunnel from the CB of the $n^+$-(In,Fe)As to the VB of the $p^+$-InAs. The kink structures of the $dI/dV$-$V$ curves are observed at the end of the tunnelling regions, marked by black arrows in Figs. 1d,e. Clearer splitting structures can be seen at the $d^2I/dV^2$-$V$ curves of devices A and B, as shown in Figs. 2a-d. Two-valley splitting structures, corresponding to the kink structures of $dI/dV - V$ curves, are clearly seen at low temperatures. The two valleys in the $d^2I/dV^2$-$V$ curves approach with increasing temperature, and merge above 45K in device A and 65K in device B, which agree well with $T_C$ of the (In,Fe)As layers in the two devices. These results strongly support our assignment that these two-valley structures correspond to the *spontaneous* spin-splitting in the CB of the (In,Fe)As layers.

To analyse these two-valley structures, we use the following fitting function, which is the sum of two Lorentzian curves and a linear offset:

$$d^2I/dV^2 = A_{minor} \frac{\Delta_{minor}/2}{(V-V_{minor})^2 + (\Delta_{minor}/2)^2} + A_{major} \frac{\Delta_{major}/2}{(V-V_{major})^2 + (\Delta_{major}/2)^2} + BV + C$$

, where $A_{minor}$ ($A_{major}$) is the magnitude, $\Delta_{minor}$ ($\Delta_{major}$) is the full width at half maximum, $V_{minor}$ ($V_{major}$) is the center bias voltage of the valley corresponding to the minority (majority) spin CB, respectively, and $BV + C$ is a linear offset. We note that the linear slope $B$ is needed only in the case of device A because of the linear offset in the vicinity of zero bias of the $d^2I/dV^2 - V$ curves. For device B, the $B$ parameter was set to be 0. The valley center's positions $V_{minor}$ and $V_{major}$, which correspond to the bottom edges of the minority and majority spin CBs of (In,Fe)As, are marked by black dots in Figs.2a,c and white dots in Figs.2b,d for each curve. From the difference between $V_{\text{minor}}$ and $V_{\text{major}}$, we estimated the spin-split energy $\Delta E$ of the (In,Fe)As CB (see Supplementary Information) and plotted as a function of temperature $T$ in Fig. 2e. One can see that large $\Delta E$ persists up to high temperatures close to $T_C$ of both devices. The error bars of $\Delta E$ are estimated by summing the standard errors of the fitting parameters $V_{\text{major}}$ and $V_{\text{minor}}$, which are smaller than 1 meV in almost all the data points. We also show in Fig. 2e two Brillouin-function fitting curves (two dotted curves) of $\Delta E$ calculated with the total angular momentum $J = 5/2$ for the $Fe^{3+}$ state, and with $\Delta E$ (at 3.5 K) and $T_C$ fitted to the experimental values. The red dotted curve was calculated with $T_C = 42$ K and $\Delta E = 32$ meV (device A, at 3.5 K), and the blue dotted curve was calculated with $T_C = 65$ K and $\Delta E = 50$ meV (device B, at 3.5 K). Both the Brillouin-function fitting curves explain quite well the temperature dependence of $\Delta E$ data in devices A and B. However, one can see that the experimental value of $\Delta E$ does not simply increase proportionally with $x$: Comparing with the $\Delta E$ and $x$ values of device A, $\Delta E$ of device B increases by 1.6 times, whereas the increase in $x$ is only 1.3 times. This deviates from the prediction of the mean-field Zener model. A much larger deviation from the mean-field Zener model is the relation between the experimental value of $\Delta E$ and $T_C$ of the same device, which will be discussed later in the Discussion section.

Next we investigate the dependence of the CB structure of (In,Fe)As on the external magnetic field $H$, applied parallel to the [100] axis in the film plane. The bottom panels of Figs. 3a-d show the $d^2I/dV^2$-$V$ curves (open circles) and their fitting curves (solid curves) of devices A and B measured at various $H$ (0 ~ 1T), at 3.5K and 50K, respectively. The top panel of each figure shows the fitting Lorentzian curves, corresponding to the majority spin (solid curves) and minority spin (dotted curves) bands, at 0 T and 1 T. The evolution of $\Delta E$ with $H$ in devices A and B are summarized in Figs. 3e and f, respectively. At 3.5 K, $\Delta E$ in both devices hardly changes with $H$, indicating a saturated value at low temperature. At 50 K, however, $\Delta E$ increases as $H$ increases in both devices. At 50 K the (In,Fe)As layer in device A ($T_C$ = 45 K) is paramagnetic, whereas that in device B ($T_C$ = 65 K) is ferromagnetic. The two spin band structures in device A, which are degenerate at 0 T, start to split and $\Delta E$ reaches 27.7 meV at 1T. This large $\Delta E$ value corresponds to a giant effective $g$-factor of 478 of (In,Fe)As, which is caused by the strong $s$-$d$ exchange interaction. We note that the measurement temperature of 50 K is close to $T_C$ of device A. In this temperature range, although global ferromagnetic order in device A disappears, local ferromagnetic order possibly still remains that may effectively enhance the $g$-factor. Measurements at temperatures much higher than $T_C$, which are required to accurately estimate the $g$-factor of paramagnetic (In,Fe)As, are difficult because of the broadening of the tunnelling spectroscopy at high temperatures. Meanwhile, $\Delta E$ in device B at 50 K slightly increases from 35 meV at 0 T to 42 meV at 1 T, which is due to the existing ferromagnetic order in the (In,Fe)As layer. It is obvious that the VB of p$^+$-InAs cannot generate this large spin splitting under a magnetic field of 1 T. Thus the $\Delta E$ - $H$ curves in Figs. 3e,f, together with the $\Delta E$ – $T$ curves in Fig. 2e, provide clear evidence that the two-valley structures in the d$^2I$/d$V^2$-$V$ curves correspond to the majority spin and minority spin CB of (In,Fe)As.

The temperature range where we can see large spin-split energy in (In,Fe)As CB is limited only by the sample's $T_C$. Therefore, if $T_C$ is increased, these results will open a way to realize FMSs with spin-splitting of the host semiconductor's band structure at high temperature, which are essential for spintronic device applications using spin-dependent band engineering. It is noteworthy that the Fe concentration $x$ (6 and 8%) and electron density $n$ (~$1 \times 10^{19}$ cm$^{-3}$) in the present (In,Fe)As samples are still far below the maximum values achieved in Mn-doped III-V FMSs (the maximum Mn doping concentration is ~20%[24] and the maximum hole density is ~$10^{21}$ cm$^{-3}$). Thus there is still much room for increasing either $x$ or $n$, which hopefully leads to higher $T_C$ in (In,Fe)As[18,19,22,23] as is commonly observed in carrier-induced FMSs. The highest $x$ that has been reported so far for (In,Fe)As is 9%[18], but this can be increased by optimizing the growth conditions or using special techniques such as delta doping[25]. On the other hand, the control of $n$ by chemical doping so far has been limited only to the use of Be or Si.[19,22] Although Be or Si atoms are doped in (In,Fe)As, $n$ is limited to at most around $1 \times 10^{19}$ cm$^{-3}$ due to their amphoteric behavior and low activation rates in InAs, especially at low growth temperature. Searching for good donors, possibly by using group VI elements or increasing $n$ by electrical gating, are intriguing methods that may increase $n$ to the order of $10^{20}$ cm$^{-3}$.

**Magnetic anisotropy of the band structure of (In,Fe)As**

To properly understand the ferromagnetism of FMSs, we need further information on the band structure of the material, such as the position of the impurity band, and the magnetic anisotropy. We thus studied tunnelling anisotropic magnetoresistance (TAMR) of (In,Fe)As. Here the external magnetic field $H$ was kept fixed at 1 T and rotated in the film plane, and the d$I$/d$V$ - $V$ curves of device A were measured at various $H$ directions with every step of 10 degrees at 3.5 K. At each direction of $H$, we noticed that the d$I$/d$V$ - $V$ curves measured at the magnetic field of 1 T and -1 T are slightly different, which is possibly due to the contribution of the Hall effect of the current flowing in the p$^+$-InAs substrate (see Supplementary Information). Since the magnetoresistance is expected to be an even function of magnetic field, we obtained the d$I$/d$V$ - $V$ curve at each direction of $H$ by averaging the two d$I$/d$V$ - $V$ curves measured at the magnetic field of 1 T and -1 T.

Figure 4a plots the change of $\frac{dI}{dV}$: $\Delta(\frac{dI}{dV}) = (\frac{dI}{dV} - \langle\frac{dI}{dV}\rangle_\phi) / \langle\frac{dI}{dV}\rangle_\phi \times 100$ (%), where $\phi$ is the magnetic field angle from the [100] axis in the clockwise direction in the film plane, and $\langle\frac{dI}{dV}\rangle_\phi$ is the $\frac{dI}{dV}$ averaged over $\phi$ at each fixed $V$. We focus on the tunnelling region ① (top panel) and the diffusion region ③ (bottom panel). Note that the color scales in the two panels are different by two orders of magnitude. No significant change was observed in the band gap region ② (not shown). One can see the TAMR data in the tunnelling region show a four-fold symmetry and another higher-order (eight-fold) symmetry, while those in the diffusion region are dominated by two-fold terms. The symmetry axis of the two-fold symmetry in the diffusion region rotates by 45 degrees (from [010] to [110]) as the bias voltage $V$ increases from 0.42 to 0.49 V. This indicates that there are at least two two-fold terms with different symmetry axes in this region.

In Fig. 4b we show the cross-sectional data (blue circles) and the fitting curves (red curves) at six bias points in Fig. 4a: Point 1 ($V$ = 50 mV) belongs to the tunnelling region, and points 2 - 6 ($V$ = 440 – 480 mV) belong to the diffusion region. In the tunnelling region, $\Delta(\frac{dI}{dV})$ was fitted by $\Delta(\frac{dI}{dV}) = C_{4[100]}\cos(4\phi) + C_{8[100]}\cos(8\phi)$, where $C_{4[100]}$ and $C_{8[100]}$ are the anisotropy constants of four-fold and eight-fold symmetry along the [100] axis, respectively. On the other hand, $\Delta(\frac{dI}{dV})$ in the diffusion region can be well fitted by $\Delta(\frac{dI}{dV}) = C_{4[100]}\cos(4\phi) + C_{2[010]}\cos[2(\phi+\frac{\pi}{2})] + C_{2[110]}\cos[2(\phi+\frac{\pi}{4})]$, where $C_{2[010]}$ and $C_{2[110]}$ are the anisotropy constants of two two-fold terms with [010] and [110] axes, respectively. The anisotropy constants estimated in the diffusion region are summarized in Fig. 4c.

TAMR originates from the change of the DOS of (In,Fe)As with changing the magnetization direction, which is caused by the spin-orbit interactions (SOI) and the *s,p-d* exchange interactions[26-28]. While in the tunnelling region electrons flow only from the (In,Fe)As CB, in the diffusion region electrons flow from the CB, VB and possibly occupied Fe-related IBs of (In,Fe)As to the empty states in the CB and VB of $p^+$-InAs, as illustrated in the inset of Fig. 4c. The different symmetries in the two regions thus come from different components of the (In,Fe)As band structure. The four-fold symmetry likely originates from the CB and VB of (In,Fe)As, which has the cubic symmetry of the zinc-blende crystal structure. The intensity $C_{4[100]}$ in the tunnelling region (0.035%) is much smaller than that in the diffusion region (~ 0.3%) because SOI in the CB is much weaker than in the VB. The existence of eight-fold symmetry is very unique to (In,Fe)As, and has been observed in the crystalline anisotropic magnetoresistance (AMR) of (In,Fe)As thin films[29]. On the other hand, the origin of the two two-fold terms ($C_{2[010]}$ and $C_{2[110]}$) are likely related to the Fe-related IBs. The bias voltages where the four-fold term ($C_{4[100]}$) and the two-fold terms ($C_{2[010]}$ and $C_{2[110]}$) appear in the diffusion region are very close, as shown in Fig. 4c, indicating the overlap of these Fe-related IBs with the CB bottom and/or the VB top of (In,Fe)As. This result is consistent with the observed position of the Fe deep levels in InAs, where Fe was doped at a very low concentration[30]. This energy overlap has been proposed to induce the large *s,p-d* exchange interactions in (In,Fe)As[18,19]. These two-fold terms are not observed in the tunnelling region, possibly because the tunnelling is forbidden by the different orbital symmetry of the Fe-related IBs and the $p^+$-

InAs VB. This indicates that the Fe-related IB is irrelevant to the spin splitting observed in the tunnelling region of the two devices A and B.

## DISCUSSIONS

The observation of the spontaneous spin splitting in the CB of (In,Fe)As provides clear evidence of the interaction between electron carriers in the CB and Fe local spins. Since both $T_C$ and the spin splitting energy $\Delta E$ in CB are obtained experimentally, (In,Fe)As can serve as a litmus test for the validity of the mean-field Zener model[1,2]. Using $\Delta E = 31.7 - 50$ meV (observed), $N_0\alpha$ is estimated to be 0.21 ~ 0.25 eV in both devices. On the other hand, $T_C$ can be calculated by the following equations:

$$T_C = \frac{S(S+1)}{12 k_B N_0} A_F (N_0\alpha)^2 x \, \rho_{3D} \tag{1}$$

$$\rho_{3D} = \frac{\sqrt{2} m^{*\frac{3}{2}}}{\pi^2 \hbar^3} \sqrt{E_F} \tag{2}$$

Here, $k_B$ is the Boltzmann constant, $A_F$ is the Fermi liquid constant, $x$ is the Fe concentration, $\rho_{3D}$ is the DOS at the Fermi level $E_F$, $\hbar$ is the reduced Planck constant, $m^*$ is the electron effective mass. Using $m^* = 0.03 \sim 0.08\, m_0$ ($m_0$ is the free electron mass)[22], $A_F = 1.2$, and $N_0\alpha = 0.21 \sim 0.25$ eV, $T_C$ is calculated to be 0.093 ~ 1.55 K, which is much lower than the observed $T_C$ of 45 - 65 K by a factor of ~ 400. Inversely, if we estimate the $N_0\alpha$ values from the experimental $T_C$ values of devices A and B and use them to calculate $\Delta E$, it should be above 1 eV, which is much larger than the observed $\Delta E = 31.7 - 50$ meV. This large discrepancy indicates that the mean-field Zener model is not applicable even to (In,Fe)As, a FMS that is free from the IB conduction picture.

Besides the case of (In,Fe)As, the validity of the mean-field Zener model in explaining the magnetic properties of other FMSs has also been discussed theoretically and experimentally[31,32,33]. The mean-field Zener model proposed a tendency of higher $T_C$ in wider-gap FMSs such as nitrides and oxides, and it concluded that the ferromagnetism in narrow-gap FMSs should be very weak[1,2]. In fact, however, while the realization of high-$T_C$ ferromagnetism in (Ga,Mn)N is still challenging, several experimental results have reported strong ferromagnetism in FMSs with narrow-gap hosts such as InAs, GaSb, and InSb. For instance, remarkably high $T_C$ (>300 K) has been recently realized in molecular beam epitaxy (MBE) grown Fe-doped FMS, (Ga,Fe)Sb[34,35,36]. Mn-doped narrow-gap FMSs grown by metal organic chemical vapor deposition (MOVPE) such as (In,Mn)As[37,38], (In,Mn)Sb[39] are other intriguing cases that show very high $T_C$. While the mechanism of these high-$T_C$ ferromagnetism is still not clearly understood, the resonance in energy of the magnetic impurity levels and the CB or VB of the host materials has been proposed as an important factor[18,19,31,33-36], which was not considered in the mean-field Zener model. Building an appropriate unified model for FMSs thus remains an unsolved theoretical challenge.

## METHODS

**Sample preparation and characterizations:** All the samples were grown by molecular beam epitaxy (MBE) on p$^+$-type InAs (001) substrates. The p$^+$-type InAs substrates were deoxidized in our MBE growth chamber at 480°C. Then, a 250 nm-thick Be doped p-type InAs buffer layer and a 5nm-thick undoped InAs buffer layer were successively grown at 460°C. The doping level of Be acceptors in the p$^+$-InAs buffer is $1 \times 10^{18}$ cm$^{-3}$. The undoped InAs buffer layer serves as a thin barrier against the diffusion of Be atoms. After cooling the substrate temperature to 236°C, a 50-nm-thick $(In_{1-x},Fe_x)As$ thin film was grown. In sample B, we doped Be in the (In,Fe)As layer at a doping level of $5\times10^{19}$ cm$^{-3}$. Although Be is a well-known acceptor when replacing the group-III site in III-V semiconductors (including InAs), we have shown that when grown at low temperature (236°C) by MBE, Be dopants mainly become double-donors in (In,Fe)As layers, probably because they favorably sit in the interstitial sites[22]. We observed bright and streaky *in situ* RHEED patterns during the MBE growth of the (In,Fe)As layers in both samples A and B, indicating good growth conditions and high crystal quality of these samples.

The magnetic properties of the (In,Fe)As layers in the two diode devices were characterized by magnetic circular dichroism and superconducting quantum interference device (SQUID) magnetometry. $T_C$ was estimated to be 45 K in A and 65 K in B, respectively. Details of the sample characterizations are given in Supplementary Information.

**Device fabrications and *I-V* measurements:** We fabricated circular-shaped mesa diodes with 200 μm in diameter using standard photolithography and Ar ion milling. A passivation layer was formed on the sample surface by spin-coating an insulating negative resist (OMR-85, ©Tokyo Ohka). A contact hole of 180 μm in diameter was opened on the top of each mesa by photolithography, and a 200 nm-thick Au electrode with 700 μm in diameter was formed on each mesa by vacuum evaporation and chemical etching. Au wires are bonded to the Au electrode and on the backside of the p$^+$-InAs substrate by In bonding contact for two-terminal transport measurements. The bias polarity was defined so that in the positive bias current flows from the p$^+$-type InAs substrate to the n$^+$-type (In,Fe)As layer (forward bias). $dI/dV$-$V$ and $d^2I/dV^2$-$V$ characteristics were numerically obtained from the *I-V* data.

**Acknowledgments**

This work is supported by Grants-in-Aid for Scientific Research including the Specially Promoted Research, the Project for Developing Innovation Systems of MEXT, Spintronics Research Network of Japan (Spin-RNJ), and the Cooperative Research Project Program of RIEC, Tohoku University. L. D. A. acknowledges the JSPS Fellowship for Young Scientists (No. 257388) and the MERIT Program. P. N. H. acknowledges Murata Science Foundation, Yazaki Foundation for Science and Technology, and Toray Science Foundation.

**Figure captions**

FIGURE 1. **a.** Device structure and transport measurement configuration of the spin-Esaki diodes used in this study. **b.** Band profiles corresponding to the case of zero bias and three regions (①: Tunnelling region, ②: Bandgap region, ③: Diffusion region) of the spin-Esaki diode. The (In,Fe)As CB is spin-split, and two red arrows represent the up and down spin bands. $E_g$ is the bandgap energy of (In,Fe)As, $E_F$ is the Fermi level, $E_n$, $E_p$ are the quasi-Fermi levels in the n$^+$-(In,Fe)As and p$^+$-InAs electrodes, respectively. Green lines in the bandgap represent gap states coming from heavily-doped impurities. **c.** Illustration of the change in the band structure of (In,Fe)As and the d$I$/d$V$ – $V$ curve of the Esaki diode at temperature above (left panel) and below (right panel) $T_C$. The spin splitting in DOS of the (In,Fe)As CB leads to the "kink" structure (pointed by two black arrows) in the d$I$/d$V$ – $V$ curve at low temperature ($T < T_C$). **d,e.** d$I$/d$V$-$V$ characteristics of devices A and B measured at 3.5 K, respectively. Blue, yellow and pink areas correspond to the tunnelling region, band-gap region, and diffusion region in each device. Blue triangles indicate the bending positions. Two black arrows point to the kink structures in each d$I$/d$V$-$V$ curve.

FIGURE 2. **a** and **c**. $d^2I/dV^2$-$V$ characteristics of devices A and B, respectively, measured at various temperatures (the vertical axes are intentionally shifted for clear vision). **b** and **d**. Color plots of the data in **a** and **c**, respectively. White and black dots in **a** - **d** mark the center positions of the valleys obtained by fitting two Lorentzian curves to the experimental $d^2I/dV^2$-$V$ characteristics. **e**. Temperature dependence of $\Delta E$ in devices A (red circles) and B (blue circles) and their Brillouin-function fitting curves (dotted red and blue curves for devices A and B, respectively).

FIGURE 3. **a – d.** Evolution of the $d^2I/dV^2$-$V$ curves with external magnetic fields. Experimental data (open circles) and their fitting curves (solid curves) are shown in the bottom panels, where white and black arrows indicate the minority and majority valley centers, respectively (The vertical axes are shifted for clear vision). The fitting curves are the sum of two Lorentzian curves, fitted for each valley. The fitting Lorentzian curves for the data at 0T (black) and 1T (red) are shown in the top panels. **a** and **b** are the results of device A at 3.5 K and 50 K, **c** and **d** are the results of device B at 3.5 K and 50K, respectively. **e** and **f.** Zeeman splitting energies $\Delta E$ at 3.5 K and 50 K of devices A (**e**) and B (**f**). The external magnetic field was applied along the [100] axis in the film plane.

FIGURE 4. **a.** Color plots of $\Delta(\frac{dI}{dV})$ in the tunnelling region (top panel) and the diffusion region (bottom panel) of device A. **b.** Cross-sectional data (blue circles) and fitting curves (red curves) at six bias points in **a**: Point 1 ($V$ = 50mV) belongs to the tunnelling region, while points 2 - 6 ($V$ = 440 – 480 mV) belong to the diffusion region. **c.** Anisotropy constants estimated in the diffusion region. The electron flows in the diffusion region are illustrated as blue arrows in the inset. Possible positions of the Fe-related IBs are also sketched as pink areas. All the data were measured at 3.5 K.

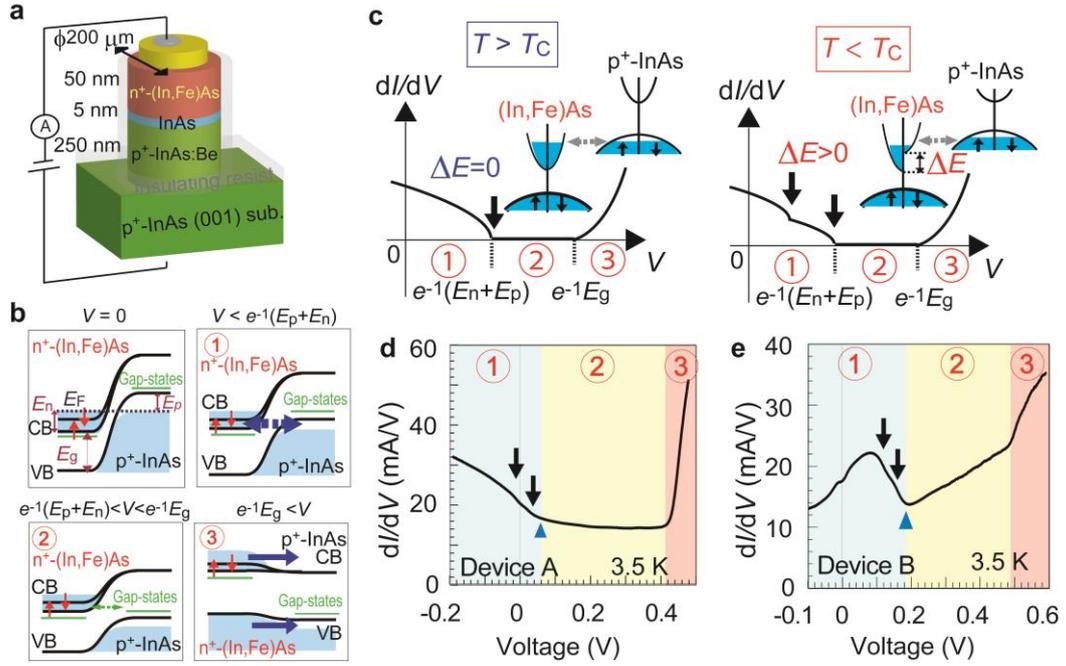

**FIGURE 1.** Anh *et al.*

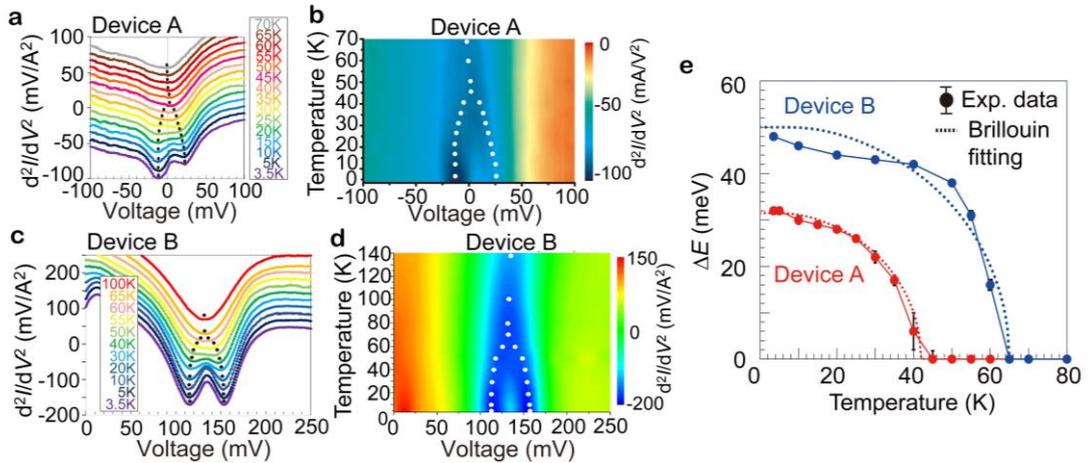

**FIGURE 2.** Anh *et al.*

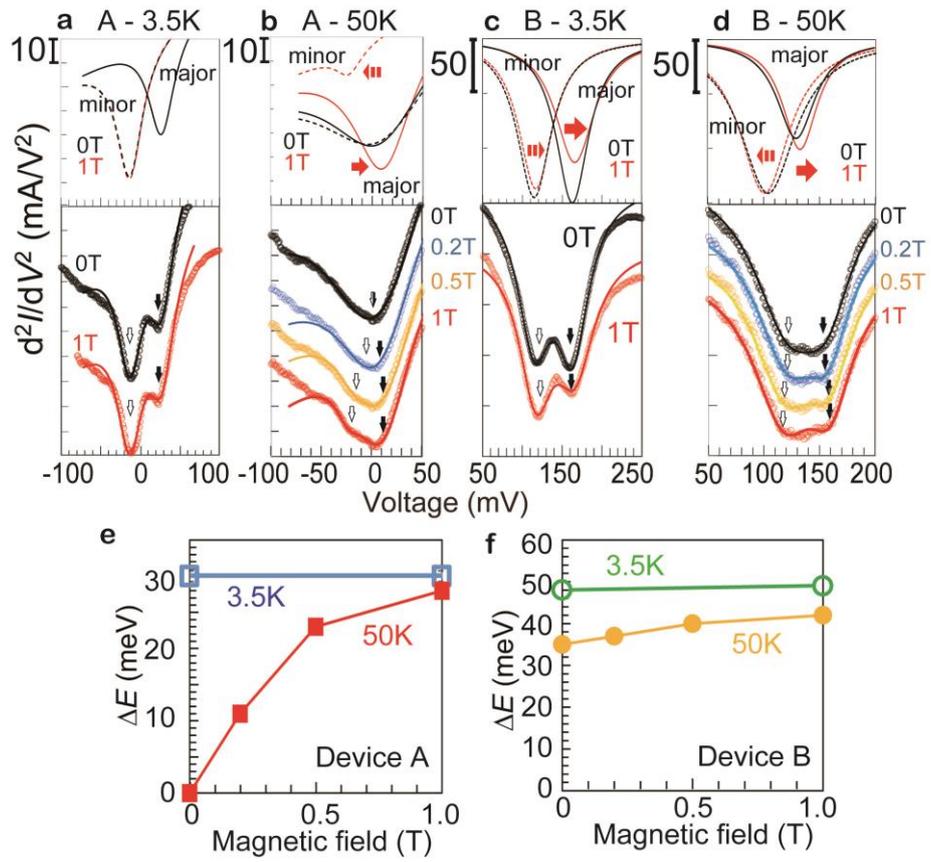

**FIGURE 3.** Anh *et al.*

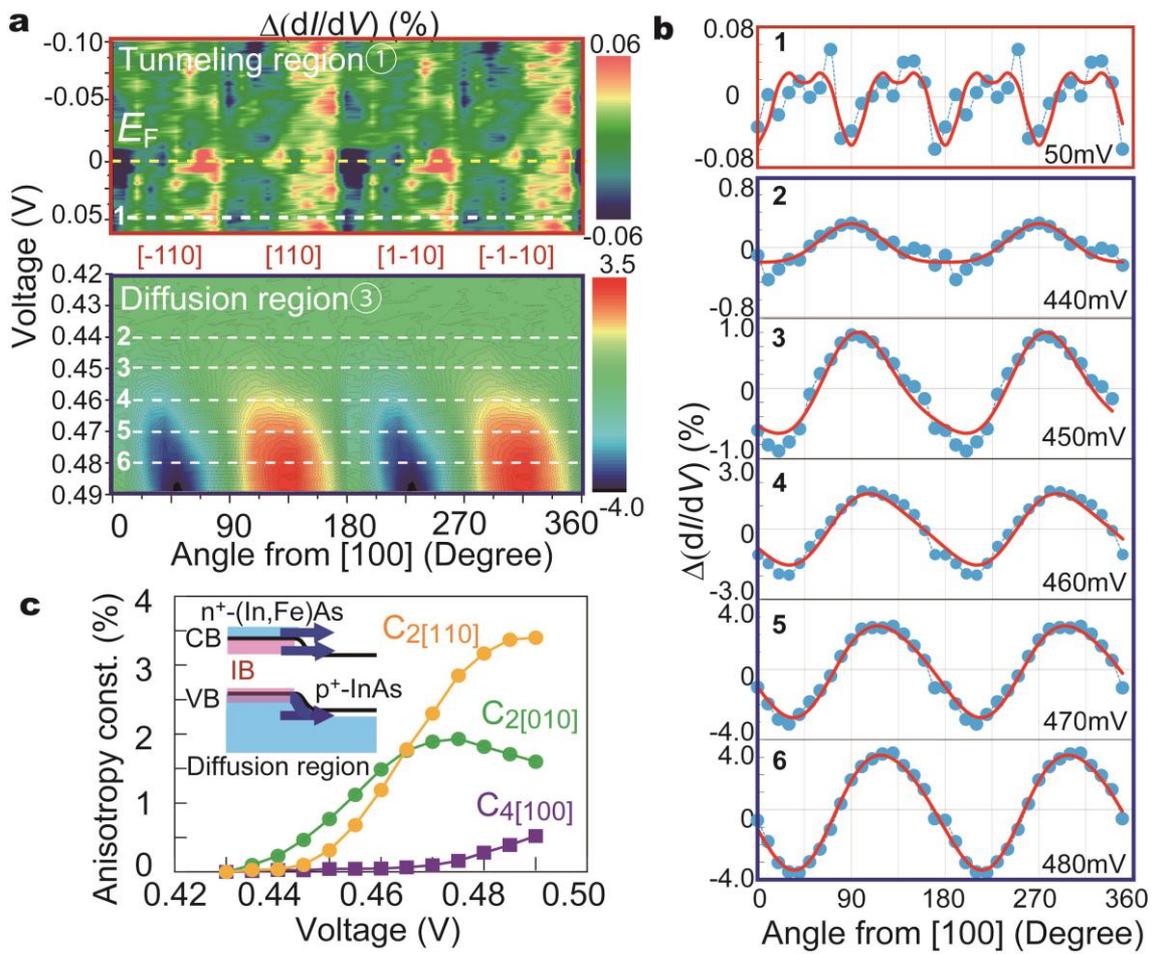

**FIGURE 4.** Anh *et al.*

# Supplementary Information

# Observation of spontaneous spin-splitting in the band structure of an n-type zinc-blende ferromagnetic semiconductor

Le Duc Anh[1,2], Pham Nam Hai[3,4], and Masaaki Tanaka[1,4]

[1]*Department of Electrical Engineering and Information Systems, The University of Tokyo, 7-3-1 Hongo, Bunkyo-ku, Tokyo 113-8656, Japan*
[2]*Institute of Engineering Innovation, Graduate School of Engineering, The University of Tokyo, 7-3-1 Hongo, Bunkyo-ku, Tokyo 113-8656, Japan*
[3]*Department of Physical Electronics, Tokyo Institute of Technology, 2-12-1 Ookayama, Meguro, Tokyo 152-0033, Japan*
[4]*Center for Spintronics Research Network (CSRN), The University of Tokyo, 7-3-1 Hongo, Bunkyo-ku, Tokyo 113-8656, Japan*

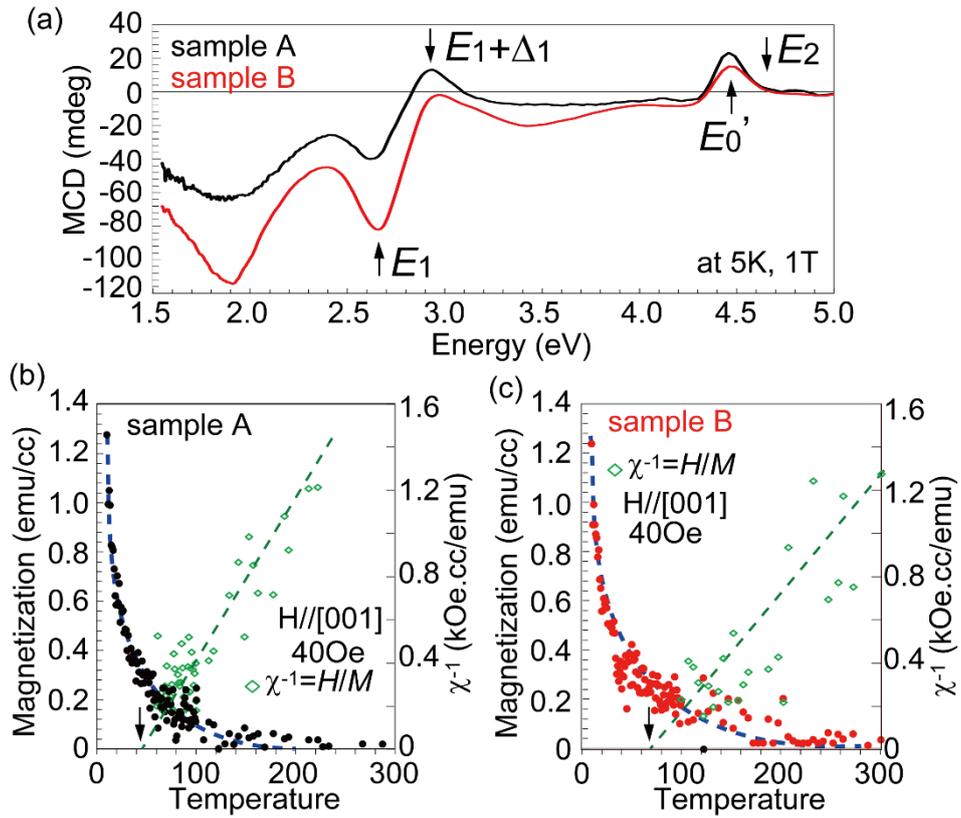

**Supplementary Figure S1:** (a) MCD spectra of sample A (black curve) and sample B (red curve) measured at 5 K, under a perpendicular magnetic field of 1 T. (b) and (c) Temperature dependence of magnetization $M$ (black or red circles, left axes) and inverse of the magnetic susceptibility $\chi$ (green open diamonds, right axes) of samples A and B, respectively, measured by SQUID. A small perpendicular magnetic field of 40 Oe was applied during the measurements. Blue dotted curves trace the $M$-$T$ curves, serving as guides to the eye. Green dotted lines are the linear approximations of the $\chi^{-1}$ data, which roughly indicate the $T_C$ values of 45 K for sample A and 65 K for sample B, as shown by black arrows.

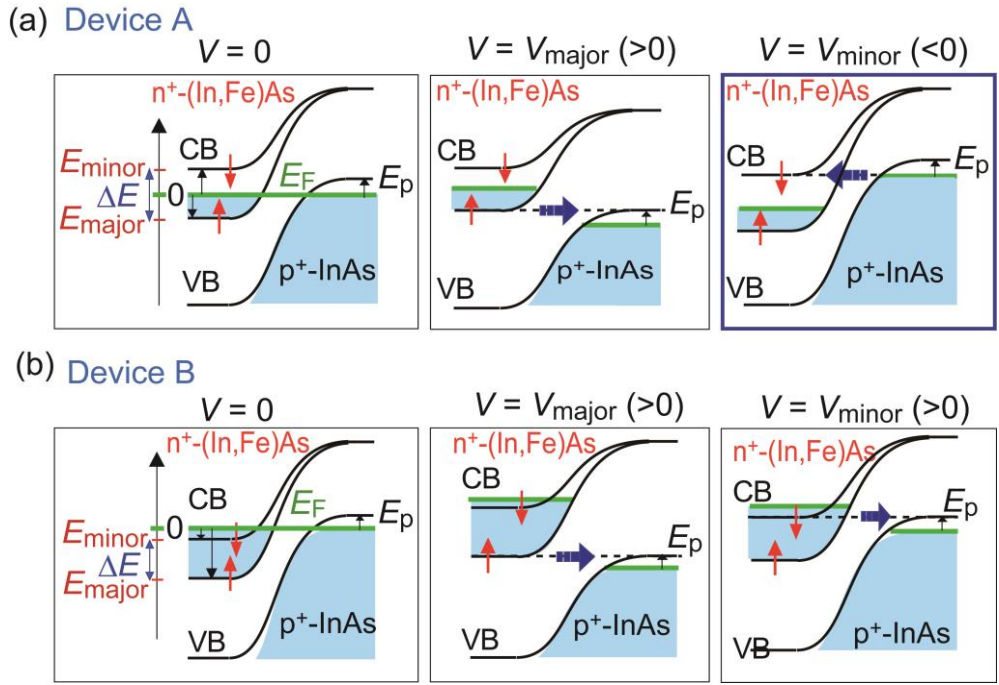

**Supplementary Figure S2.** Band profiles of the pn junctions in the two devices A (a) and B (b), respectively, at bias voltage $V = 0$, $V_{major}$, and $V_{minor}$, all at low temperature (ferromagnetic state). At $V = 0$, with the Fermi energy at the zero point, the energy levels of the majority and minority spin CB bottom edges of (In,Fe)As and of the VB top of $p^+$-InAs are denoted as $E_{major}$, $E_{minor}$, and $E_p$, respectively. The blue arrows show the tunnelling directions of electrons at non-zero bias $V$. In device A, the tunnelling directions of electrons at $V = V_{major}$ and at $V = V_{minor}$ are opposite to each other.

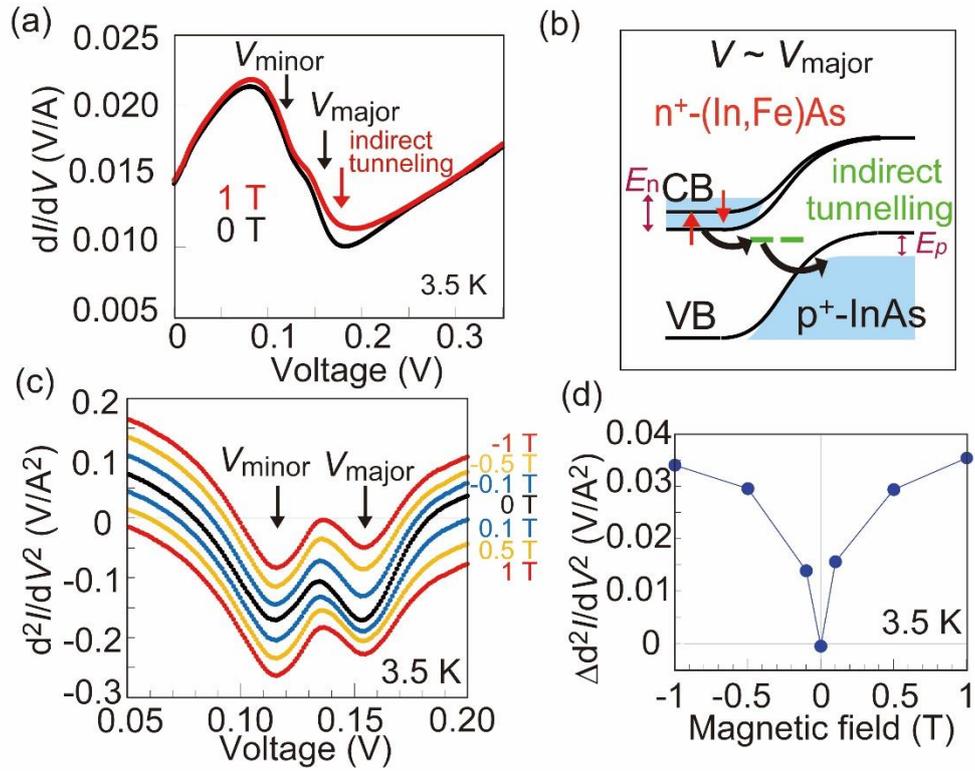

**Supplementary Figure S3.** (a) d$I$/d$V$ – $V$ curves of device B, measured at 3.5 K without and with a magnetic field $H$ of 1 T applied in the film plane. At 1 T, d$I$/d$V$ is increased at the end of the tunnelling region ($V \sim V_{major}$), which causes the shallowing of the majority spin valley in the d$^2I$/d$V^2$ – $V$ curves at 1 T observed in the upper panel of Fig. 3c in the main manuscript. (b) Schematic energy diagram of the p-n junction at $V \sim V_{major}$. The enhancement of d$I$/d$V$ is probably due to indirect tunnelling processes at the interface. (c) d$^2I$/d$V^2$ – $V$ curves of device B at 3.5 K, measured under various magnetic fields $H$ from -1 T to 1 T applied in the film plane. (d) Difference in the d$^2I$/d$V^2$ values of (c) at the majority and minority spin's valley centers, $\Delta$d$^2I$/d$V^2$ = d$^2I$/d$V^2$($V_{major}$) - d$^2I$/d$V^2$($V_{minor}$), as an indicator of the asymmetry between the two valleys under different $H$.

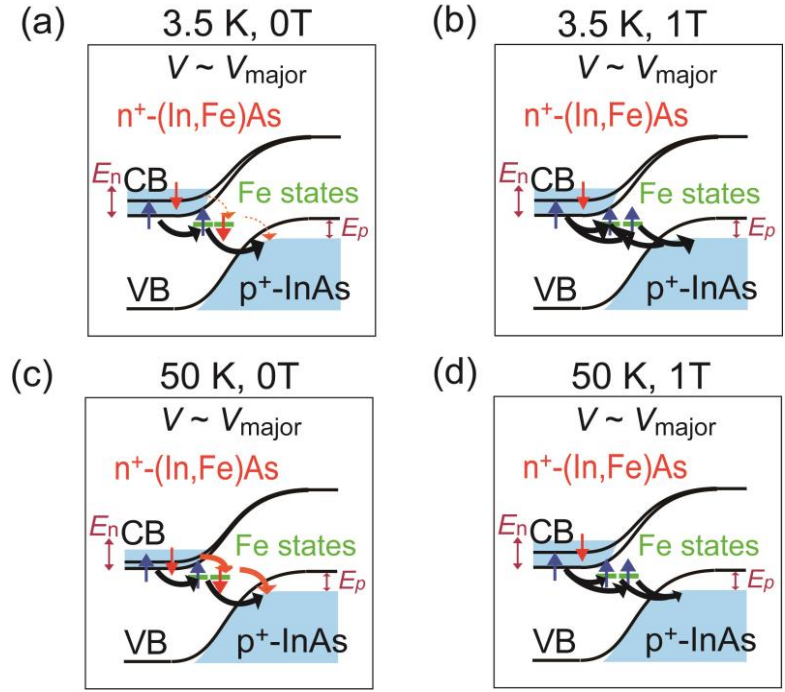

**Supplementary Figure S4.** Schematic energy diagram of the gap-state assisted tunnelling through paramagnetic Fe states (green lines) in device B. (a) and (b) are the situations at 3.5 K, without and with $H$, whereas (c) and (d) are the situations at 50 K, without and with $H$, respectively. At 3.5 K and 0 T (panel a) the gap-state assisted tunnelling is contributed mainly from the majority CB (blue arrow), whose energy is close to that of the Fe gap states. At 50 K and 0 T (panel c) both of the majority spin (blue arrow) and minority spin (red arrow) CBs contribute to the Fe gap-state assisted tunnelling. At 1 T (panels b and d), the spin angular momentums of the Fe gap states are aligned with the majority spin in the CB of (In,Fe)As, thus the tunnelling from the minority spin CB is prohibited, while the probability from the majority spin CB is enhanced.

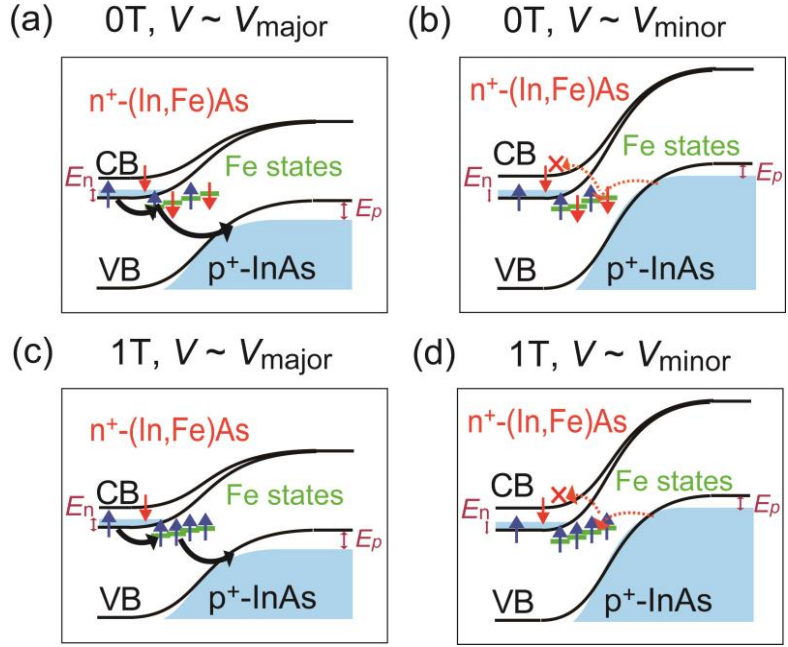

**Supplementary Figure S5.** Schematic energy diagrams of the gap-state assisted tunnelling through paramagnetic Fe states (green lines) in device A at 3.5 K. (a) and (b) are the situations at 0 T, at $V_{major}$ and $V_{minor}$, whereas (c) and (d) are the situations at 1 T, at $V_{major}$ and $V_{minor}$, respectively. The Fe gap state assisted tunnelling occurs at $V = V_{major}$, which is contributed by electrons in the majority spin CB, but is zero at $V = V_{minor}$. Due to the small electron density $n$ of (In,Fe)As in device A (~$1\times10^{18}$ cm$^{-3}$), the depletion layer of the p-n junction extends more into the (In,Fe)As side, which increases the number of the Fe gap states and the indirect tunnelling current. At 1T (panel c, d) the spin angular momentums of the Fe gap states are aligned with the majority spin in the CB of (In,Fe)As. However due to the small $n$ of the (In,Fe)As layer, the indirect tunnelling current from the majority spin CB is almost unchanged.

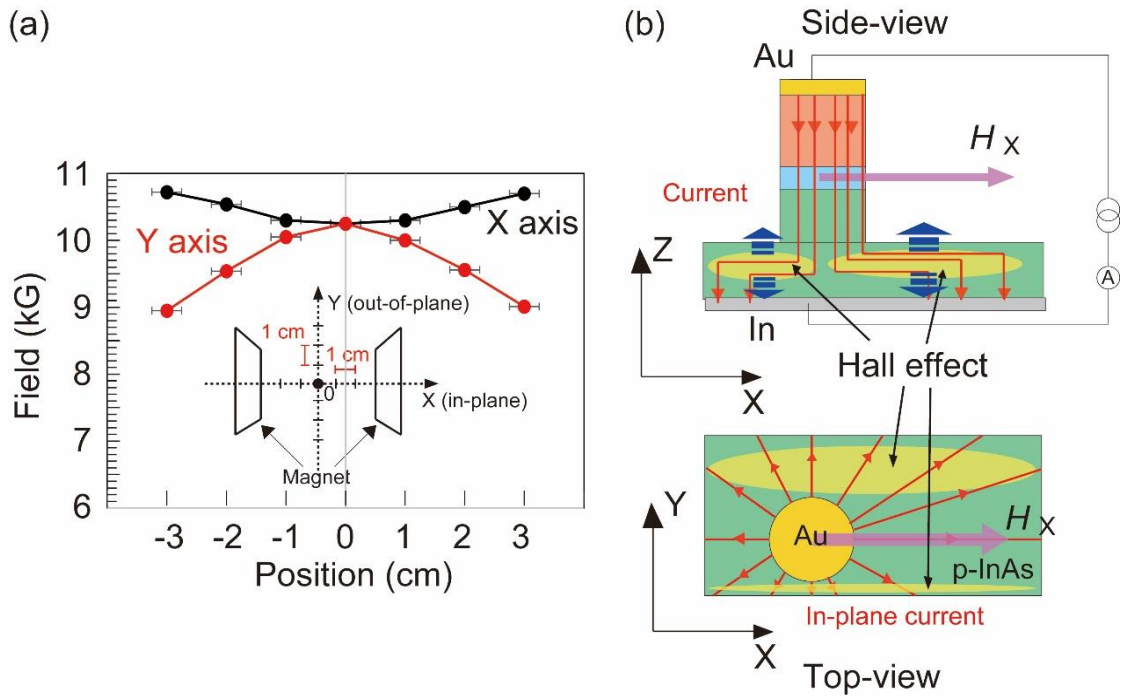

**Supplementary Figure S6.** (a) Magnetic field measured in the sample space between two poles of the electromagnet by a Gaussmeter, which confirmed the even-function symmetry of the magnetic field around the center position (inset). (b) Schematic sketch of the diode device placed in the magnetic field $H_X$, which is along the X direction. Red lines illustrate the current paths, which are mainly in the Z direction. However, in the $p^+$-InAs substrate (yellow areas) currents flow in the X-Y plane. The magnetic field $H_X$ and the currents with Y-direction components induce Hall voltages in the Z direction. These Hall voltages, which are odd functions of the magnetic field, are not canceled out due to different path lengths of the currents in the substrate, thus can be detected by the top and bottom electrodes.

## Supplementary Note 1:
## Characterization of the magnetic properties

The magnetic properties of the two samples A and B were characterized by magnetic circular dichroism (MCD) in a reflection configuration and superconducting quantum interference device (SQUID) measurements.

Supplementary Figure S1(a) shows MCD spectra of samples A (black curve) and B (red curve), measured at 5 K and under 1 T applied perpendicular to the film plane. The MCD spectra of both samples show peaks corresponding to the optical transition at the critical points of InAs zinc blende (ZB) band structure: $E_1$ (2.61 eV), $E_1+\Delta_1$ (2.88 eV), $E_0$' (4.39 eV), and $E_2$ (4.74 eV)[1]. The MCD peak at lower energy (~2eV) is usually observed in bulk-like thick (In,Fe)As samples, although its origin is not clear at this stage. The MCD spectra confirmed the ZB-type crystal structure and band structure of (In,Fe)As in the two samples. One can also see that the MCD intensity of sample B (Fe content: 8%) is twice as large as that of sample A (Fe content: 6%), reflecting the different magnetization in the two samples.

Supplementary Figures S1(b) and (c) show the temperature dependence of magnetization (*M-T* curves) of the two samples A and B, respectively, measured by SQUID when increasing temperature from 10 K to 300 K. These are field cooling (FC) curves; before the *M-T* measurements, the samples were cooled to 10 K with a magnetic field of 1 T applied perpendicular to the film plane. Note that because of the small remanent magnetization in both samples, a weak perpendicular magnetic field of 40 Oe was applied during these *M-T* measurements which caused a tail up to temperatures above $T_C$. One can see that the magnetization of sample B persisted at high temperatures (100 K < *T* < 200 K), while that of sample A decreased quickly and was almost unchanged above 100 K. The inverse of the magnetic susceptibility $\chi^{-1} = H/M$ [at $H =$ 40 Oe, *T* > 100 K, green open diamonds] follows the Curie-Weiss law $\chi = C/(T-T_C)$ [green dotted line, $T_C$ is the Curie temperature, *C* is the Curie constant], allowing us to estimate $T_C$ of the two samples. Although the $\chi^{-1}$ data points are fluctuating due to the low S/N ratio at high *T*, we can roughly estimate the $T_C$ to be 45 K for sample A and 65 K for sample B, respectively.

## Supplementary Note 2:
## Estimation of the spin split energy $\Delta E$ of (In,Fe)As from the center positions of the two-valley structures in the $d^2I/dV^2 - V$ curves

The two spin-Esaki diode devices (A and B) studied in this work differ in Fe concentration (6% and 8%, respectively) and electron density (due to co-doping of Be double donors in the (In,Fe)As layer in device B). Both of the two diode devices show two-valley structures in the $d^2I/dV^2 - V$ curves (Figs. 2a-d and Figs. 3a-d in the main manuscript), corresponding to the splitting of the majority spin conduction band (CB) and minority spin CB of (In,Fe)As. For the two-valley structures, we fitted the sum of two Lorentzian curves to determine the valley center positions of the majority and minority spin CBs ($V_{major}$ and $V_{minor}$, respectively) (see main manuscript). The spin split energy $\Delta E$ of the (In,Fe)As layers was estimated by the difference between $V_{major}$ and $V_{minor}$ as follows.

Supplementary Figures S2a and b show the band profiles of the p-n junctions in the two devices A and B, respectively, at low temperature (ferromagnetic state) and bias voltage $V = 0$, $V_{major}$, and $V_{minor}$. In these figures, we set the Fermi energy at the zero point, and denote the energy levels of the majority and minority spin CB bottom edges of (In,Fe)As and the valence band (VB) top of p$^+$-InAs as $E_{major}$, $E_{minor}$, and $E_p$, respectively. The spin split energy $\Delta E$ of (In,Fe)As is therefore given by $E_{minor} - E_{major}$.

In device A, because $V_{major}$ is positive whereas $V_{minor}$ is negative (data shown in Figs. 2a,b of the main manuscript), the Fermi level $E_F$ lies above the band edge of the majority spin CB and below that of the minority spin CB ("half-metallic" band structure, i.e. $E_{major} < 0$, $E_{minor} > 0$), as illustrated in Supplementary Figures S2a. At $V_{major}$, the band edge of majority spin CB of (In,Fe)As is aligned with the top of the valence band (VB) of p$^+$-InAs, thus $eV_{major} = -E_{major} + E_p$. However, at $V_{minor}$, which is negative, the band edge of the minority spin CB of (In,Fe)As is aligned with the quasi-Fermi level of the p$^+$-InAs, thus $eV_{minor} = -E_{minor}$. Therefore, we have the following relation:

$$e(V_{major} - V_{minor}) = E_{minor} - E_{major} + E_p = \Delta E + E_p \qquad (S1)$$

This means that the difference between $V_{major}$ and $V_{minor}$ overestimates the spin split energy $\Delta E$ of (In,Fe)As CB in device A by $E_p$. To correct this overestimation, we have subtracted $e(V_{major} - V_{minor})$ of device A by $E_p$, which is estimated to be 8.3 meV from the Be doping concentration of $1 \times 10^{18}$ cm$^{-3}$.

On the other hand, in device B, because both $V_{\text{major}}$ and $V_{\text{minor}}$ are positive (data shown in Figs. 2c,d of the main manuscript), the Fermi level $E_F$ in (In,Fe)As lies above both the majority and minority spin CB bottom edges as illustrated in Supplementary Figures S2b (thus, $E_{\text{major}} < 0$ and $E_{\text{minor}} < 0$). At $V_{\text{major}}$ ($V_{\text{minor}}$), the band edge of the majority (minority) spin CB of (In,Fe)As is aligned with the <u>top of the VB</u> of p$^+$-InAs. Therefore we have the following relations:

$$eV_{\text{major}} = -E_{\text{major}} + E_{\text{p}} \tag{S2}$$

$$eV_{\text{minor}} = -E_{\text{minor}} + E_{\text{p}} \tag{S3}$$

$$e(V_{\text{major}} - V_{\text{minor}}) = E_{\text{minor}} - E_{\text{major}} = \Delta E \tag{S4}$$

Thus the difference between $V_{\text{major}}$ and $V_{\text{minor}}$ corresponds exactly to the spin split energy $\Delta E$ of (In,Fe)As CB in device B.

## Supplementary Note 3:
## Behavior of the majority and minority spin's valleys in the d$^2$I/dV$^2$ – V curves of the two devices under various magnetic fields

From the upper panels of Figs. 3a, c, and d (the minority and majority spin Lorentzian curves of each experimental data), one can see that in Figs. 3a and d, the majority spin valley is smaller (shallower) than the minority spin valley at zero magnetic field, whereas in Fig. 3c the majority spin valley became shallower with applying $H$. In the following, we explain possible reasons for this complicated behavior.

First, we show in Supplementary Figures S3a the d$I$/d$V$ – $V$ curves of device B at 3.5 K and under 0 T (black) and 1 T (red). One can see that the shallowing of the majority spin valley in the d$^2$I/dV$^2$ – V curve is caused by the increase of the d$I$/d$V$ <u>after the end of the direct tunnelling region ($V \sim V_{\text{major}}$)</u> of the Esaki diode (indicated by the red arrow in Supplementary Figures S3a). As illustrated in Supplementary Figures S3b, at the end of the tunnelling region ($V \sim V_{\text{major}}$), the (In,Fe)As conduction band (CB) bottom (majority spin CB bottom) is lifted to the same energy as the p$^+$-InAs valence band (VB) top, and direct tunnelling from CB to VB is suppressed. Therefore, the increase of the d$I$/d$V$ after the end of the tunnelling region reflects the tunnelling conductance due to other indirect tunnelling processes, such as magnon-assisted tunnelling, phonon-assisted tunnelling, gap-state assisted tunneling, or their combinations.

In Supplementary Figures S3c, we plot d$^2$I/dV$^2$ – V curves of device B, measured at 3.5 K under various magnetic fields $H$ from -1 T to 1 T applied in the film plane (the data at 0 T and 1 T are the same as those plotted in Fig. 3c). To show the dependence of the asymmetry between the majority and minority spin's valleys on the magnetic field $H$, we plot in Supplementary Figures S3d the difference in the $d^2I/dV^2$ values at the majority and minority spin's valley center, $\Delta d^2I/dV^2 = d^2I/dV^2(V_{major}) - d^2I/dV^2(V_{minor})$, as a function of the magnetic field $H$. We see that $\Delta d^2I/dV^2 - H$ shows the same nonlinear behavior under positive and negative $H$. This result indicates that the indirect tunnelling process in device B is magnetic-field dependent.

Here, we propose a scenario of *gap-state assisted tunnelling through paramagnetic Fe-induced gap states* at the interface of the p-n junction: At the interface or in the depletion region of the p-n junction, some Fe gap states can exist due to diffusion of Fe atoms from the (In,Fe)As electrode. The energy levels of these paramagnetic Fe-induced states are close to the CB bottom of (In,Fe)As [2]. Thus, electrons at the CB bottom of (In,Fe)As can indirectly tunnel to the VB top of p$^+$-InAs through these paramagnetic Fe gap states after the end of the direct tunneling region.

Supplementary Figures S4 and S5 illustrate the schematic energy diagrams of the paramagnetic Fe gap states in devices A and B, respectively, at different temperatures and magnetic fields. Using these diagrams, we will explain the behavior of the two spin valleys in the $d^2I/dV^2 - V$ curves in Figs. R4a, c, and d, as follows.

**Behavior of the $d^2I/dV^2 - V$ curves in Fig. 3c (device B, 3.5 K)**

At 3.5 K and 0 T (Supplementary Figures S4a), the Fe gap-state assisted tunnelling occurs mainly from the majority spin CB, whose energy is close to that of the Fe gap states, to the p$^+$InAs VB through the paramagnetic Fe gap states that have the same spin magnetic moment direction (blue arrows in Supplementary Figures S4a).

At 3.5 K and 1 T (Supplementary Figures S4b), however, there are more Fe gap states whose magnetic moments aligned with the majority spins in the CB of (In,Fe)As. Thus, indirect tunnelling from the majority spin CB is enhanced. This explains the increase of the $dI/dV$ after the end of the direct tunneling region (Supplementary Figures S3a) and the shallowing of the majority spin valley of the $d^2I/dV^2 - V$ curves (Fig. 3c) when $H$ was applied.

**Behavior of the $d^2I/dV^2 - V$ curves in Fig. 3d (device B, 50 K)**

In the upper panel of Fig. 3d, the majority spin valley is shallower than the minority spin valley at 0 T, and becomes slightly deeper when applying $H$, although it is still shallower than the minority spin valley. We can explain this behavior as follows. At 50 K, electrons from both the majority spin (blue arrow) and minority spin (red arrow) CBs can tunnel through the paramagnetic Fe gap states and contribute to the indirect tunnelling current, as shown in Supplementary Figures S4c. This is possible because of the smaller spin split energy of (In,Fe)As CB and phonon-assisted processes existing at 50 K. This explains why the majority spin valley at 50 K (Fig. 3d) is shallower than that at 3.5 K (Fig. 3c) at 0 T. However, when $H$ was applied, more magnetic moments of the paramagnetic Fe states are aligned with the majority spin in the CB of (In,Fe)As, and the indirect tunnelling from the minority spin CB is partly suppressed, as shown in Supplementary Figures S4d. Therefore the total indirect tunnelling current decreases, which explains why the majority spin valley of the $d^2I/dV^2 - V$ curve at 50 K in devices B becomes less shallow (slightly deeper) with applying $H$ as seen in Fig. 3d.

## Behavior of the $d^2I/dV^2 - V$ curves in Fig. 3a (device A, 3.5 K)

In the upper panel of Fig. 3a in the main manuscript, the majority spin valley is shallower than the minority spin valley even at 0 T, and the two spin valleys change very little with $H$. Supplementary Figures S5 shows the schematic energy diagrams of the gap-state assisted tunnelling through paramagnetic Fe states (green lines) in device A at bias voltages $V = V_{major}$ and $V = V_{minor}$. In device A, because the Fermi level of (In,Fe)As lies above the bottom of the majority spin CB but below that of the minority spin CB, the tunneling direction of electrons at $V = V_{minor}$ is opposite to that at $V = V_{major}$, as illustrated in Supplementary Figures S5 (see also Supplementary Figures S2). At 0 T and $V = V_{major}$ (Supplementary Figures S5a) the Fe gap-state assisted tunnelling current is contributed only by electrons in the majority spin CB (blue arrow) of (In,Fe)As, because the minority spin CB (red arrow) is empty. Meanwhile, at $V = V_{minor}$ (Supplementary Figures S5b,d) the minority spin electrons in the VB of $p^+$-InAs, however, cannot tunnel into the minority spin CB of (In,Fe)As through the Fe gap states because the energy levels of the Fe gap states are lower than the minority CB bottom edge. Therefore the Fe gap-state assisted tunnelling current at $V = V_{minor}$ is zero. This difference of the Fe gap-state assisted tunnelling currents in the cases of $V = V_{major}$ and $V = V_{minor}$ explains why the majority spin valley is shallower than the minority spin valley at 0 T. We also note that because the electron density $n$ of (In,Fe)As is lower (~$1 \times 10^{18}$ cm$^{-3}$) than device B, the depletion layer of the p-n junction extends more into the (In,Fe)As side. Due to the lack of carriers inside the depletion region of (In,Fe)As, more Fe atoms act as paramagnetic Fe states, which increases the number of the Fe gap states. This situation further enhances the Fe gap-state assisted tunnelling current at $V = V_{major}$ in device A in comparison with that of device B.

When applying $H = 1$ T (Supplementary Figures S5c), the number of majority spin Fe gap states increases. However, due to the small electron density $n$ in the CB of (In,Fe)As layer, an increase in the number of majority spin Fe gap states (which is already quite large at 0 T) does not yield any large effect. This is why the gap-state assisted tunnelling current shows almost no change. Besides, other magnetic-field-independent mechanism (ex. phonon-assisted tunnelling) may be dominant in device A.

## Supplementary Note 4:
## Origin of the asymmetry between the d$I$/d$V$ – $V$ curves of device A measured at +1T and -1 T

In the TAMR measurements, the external magnetic field $H$ was kept fixed at 1 T and rotated in the film plane, and the d$I$/d$V$-$V$ curves of device A were measured at various $H$ directions with every step of 10 degrees at 3.5 K. At each direction of $H$, we noticed that the d$I$/d$V$-$V$ curves measured at the magnetic field of 1 T and -1 T are slightly different. The difference in the d$I$/d$V$ – $V$ curves of 1T and -1T is caused by an *odd* function contribution of the magnetic field $H$, due to the Hall effect occurring in the p$^+$-InAs substrate, as explained below.

The devices were placed meticulously in the center position of the space between the two poles of our electromagnet (misalignment, if any, should be in millimeter order). In principle, with the center position as the origin, the distribution of magnetic field in this sample space is an even function of the position. Supplementary Figures S6a show the position dependence of $H$ in the sample space of our electromagnet measured by a Gaussmeter, which confirmed the even-function symmetry of the magnetic field. Therefore misalignment of the sample from the center position, if any, cannot generate a response that is an odd function of $H$.

In Supplementary Figures S6b, we show a general situation when the sample is placed in the X-Y plane under a magnetic field $H$ applied in the X direction ($H_X$). The currents (red lines) flow through the mesa diode in the Z direction. In the thick p$^+$-InAs substrate, however, the currents can flow in the X-Y plane (yellow areas in Supplementary Figures S6b). Because the top mesa is not located exactly at the center of the substrate, different path lengths are expected for currents flowing in different directions in the X-Y plane (see the top-view in Supplementary Figures S6b). In these yellow areas, the magnetic field $H_X$ and the currents in the Y-direction induce Hall voltages in the Z direction. These Hall voltages are not canceled out due to different path lengths of the currents in the substrate, thus can be detected by the top and bottom electrodes. These Hall voltages are odd functions of the magnetic field and thus have opposite signs at +1T and -1T. We think that this is the main origin of the difference in the d$I$/d$V$ – $V$ data under +1T and -1T.